\RequirePackage{lineno}
\documentclass[
prl,
reprint,
twocolumn%
,amssymb, amsmath, nobibnotes, aps, showpacs, nofootinbib, superscriptaddress]{revtex4-1}

\usepackage[colorlinks=true,linkcolor=blue,citecolor=red]{hyperref}%
\usepackage{soul} \setstcolor{red}\setulcolor{blue}\sethlcolor{yellow}\setul{}{0.3ex}
\usepackage{graphicx}
\usepackage{float}
\graphicspath{{./}}

\newcommand{\gray}{$\gamma$-ray}
\newcommand{\fermi}{{\it Fermi}}
\newcommand{\fermilat}{{\it Fermi}-LAT}

\soulregister{\gray}{0}

\begin{document}
%\linenumbers

\title{Inferred cosmic-ray spectrum from \fermilat{} \gray{} observations of the Earth's limb}

%Author list updated Wednesday 7 Feb 2014 00:10 PST
\author{M.~Ackermann}
\affiliation{Deutsches Elektronen Synchrotron DESY, D-15738 Zeuthen, Germany}
\author{M.~Ajello}
\affiliation{Space Sciences Laboratory, 7 Gauss Way, University of California, Berkeley, CA 
94720-7450, USA}
\author{A.~Albert}
\affiliation{W. W. Hansen Experimental Physics Laboratory, Kavli Institute for Particle 
Astrophysics and Cosmology, Department of Physics and SLAC National Accelerator Laboratory, 
Stanford University, Stanford, CA 94305, USA}
\author{A.~Allafort}
\affiliation{W. W. Hansen Experimental Physics Laboratory, Kavli Institute for Particle 
Astrophysics and Cosmology, Department of Physics and SLAC National Accelerator Laboratory, 
Stanford University, Stanford, CA 94305, USA}
\author{L.~Baldini}
\affiliation{Istituto Nazionale di Fisica Nucleare, Sezione di Pisa, I-56127 Pisa, Italy}
\author{G.~Barbiellini}
\affiliation{Istituto Nazionale di Fisica Nucleare, Sezione di Trieste, I-34127 Trieste, Italy}
\affiliation{Dipartimento di Fisica, Universit\`a di Trieste, I-34127 Trieste, Italy}
\author{D.~Bastieri}
\affiliation{Istituto Nazionale di Fisica Nucleare, Sezione di Padova, I-35131 Padova, Italy}
\affiliation{Dipartimento di Fisica e Astronomia ``G. Galilei'', Universit\`a di Padova, 
I-35131 Padova, Italy}
\author{K.~Bechtol}
\affiliation{W. W. Hansen Experimental Physics Laboratory, Kavli Institute for Particle 
Astrophysics and Cosmology, Department of Physics and SLAC National Accelerator Laboratory, 
Stanford University, Stanford, CA 94305, USA}
\author{R.~Bellazzini}
\affiliation{Istituto Nazionale di Fisica Nucleare, Sezione di Pisa, I-56127 Pisa, Italy}
\author{R.~D.~Blandford}
\affiliation{W. W. Hansen Experimental Physics Laboratory, Kavli Institute for Particle 
Astrophysics and Cosmology, Department of Physics and SLAC National Accelerator Laboratory, 
Stanford University, Stanford, CA 94305, USA}
% ...

\collaboration{\fermilat{} Collaboration}
\noaffiliation

%%%%%%%%%%%%%%%%%%%%%%%%%%%%%%%%%%%%%%%%%%%%%
\begin{abstract}
  Recent accurate measurements of cosmic-ray (CR) species
  by ATIC-2, CREAM, and PAMELA reveal an unexpected hardening in
  the proton and He spectra above a few hundred GeV, a gradual softening
  of the spectra just below a few hundred GeV, and a harder spectrum of
  He compared to that of protons. These newly-discovered features
  may offer a clue to the origin of high-energy CRs. We use the
  \fermi{} Large Area Telescope observations of the \gray{} emission
  from the Earth's limb for an indirect measurement of the local spectrum
  of CR protons in the energy range $\sim 90$~GeV--6~TeV (derived from
  a photon energy range 15~GeV--1~TeV). Our analysis shows that 
  single power law and broken power law spectra fit the data equally
  well and yield a proton spectrum with index $2.68 \pm 0.04$ and
  $2.61 \pm 0.08$ above $\sim 200$~GeV, respectively.
\end{abstract}

%%%%%%%%%%%%%%%%%%%%%%%%%%%%%%%%%%%%%%%%%%%%%
% insert suggested PACS numbers in braces on next line
% insert suggested keywords - APS authors don't need to do this
%\keywords{}

\date{\today}%
\maketitle

%%%%%%%%%%%%%%%%%%%%%%%%%%%%%%%%%%%%%%%%%%%%%
\emph{Introduction.}
The spectrum of CRs has offered few clues to its origin so far. The
generally accepted features are at very-high and ultra-high
energies (see, e.g., Figure~1 in \cite{Swordy2001}): the so-called
``knee'' at a few thousand TeV \cite{knee1958,Haungs2003}, the
second ``knee'' at $\sim 10^{6}$~TeV, the ``ankle'' at higher
energies \cite{Abbasi2005}, and a spectral steepening above $10^{8}$~TeV
\cite{Abbasi2009,AUGER2010}. It is
believed that CRs below the second knee are Galactic, while
extragalactic CRs dominate at higher energies
\citep[][]{Berezinskii1990, SMP2007}.

The data recently collected by three experiments, ATIC-2
\cite{ATIC2008elements,Panov2009}, CREAM \cite{CREAM2010break,
CREAM2011pHe}, and PAMELA \cite{PAMELA2011}, indicate a new
feature at relatively low energy: a break (or hardening)
of CR proton and He spectra at $\sim 240$~GV in rigidity.
PAMELA claims to detect the break at 95\% confidence level
for both species. Below the break, PAMELA data
agree very well with the earlier data from AMS-01 \cite{AMS2000} and BESS
\cite{BESS2004}. Above the break, ATIC-2 results agree well with those
of CREAM, smoothly connecting to highest energy points from PAMELA.
The change in the spectral indices for both protons and He is $\sim 0.2$.

However, the break itself is observed only by PAMELA
near its high-energy limit. Much evidence of this newly discovered break 
or flattening comes from a combination of data by several different experiments,
which may be subject to cross-calibration errors. A verification of
this new feature requires an independent confirmation, preferably with
a single instrument. Meanwhile, recent preliminary AMS-02 
results\footnote{http://www.ams02.org/wp-content/uploads/2013/07/Proton\_2.jpg} 
do not show any feature in the proton and He spectra
up to $\sim 2$~TeV and also seem to contradict
ATIC-2 and CREAM results. In this paper, we demonstrate
that such a measurement can also be done indirectly through
observation of the CR-induced \gray{} emission from the Earth's atmosphere.

Atmospheric \gray{} emission is mainly the result of hadronic CR cascades:
CRs entering the atmosphere near grazing incidence produce showers
that develop in the forward direction, resulting in
a very bright \gray{} signal from the Earth's limb as seen from orbit.
The \gray{} spectrum from CR interactions at the very top of
the atmosphere depends only on the inclusive \gray{}
production cross section and the spectrum of CR particles.
If the cross section is known, the shape of the local CR spectrum
can be recovered from the \gray{} spectrum. However, this method
can only measure the total spectrum of CRs. To deduce the spectrum of
protons, the most abundant component of CRs, one has to assume a
spectrum of He, the second most abundant component. The contribution
of the latter to the total \gray{} emission is $\sim 10$--20\%,
depending on the energy. Therefore, accurate modeling of the contribution
from He interactions is not very critical, and heavier nuclei can be neglected.

Observations of Galactic diffuse \gray{} emission (GDE) have
provided valuable information about CR spectra in distant locations
\cite{2000ApJ...537..763S,2004ApJ...613..962S,2008ApJ...682..400P,
2009PhRvL.103y1101A,2012ApJ...750....3A} and in the local interstellar medium
\cite{2009ApJ...703.1249A,2010ApJ...710..133A}.  Similarly,
observations of the Earth's limb \gray{} emission can be used
to deduce the CR spectrum near the Earth. In contrast with the
GDE, the contribution from the inverse Compton scattering of CR
electrons to the Earth's limb emission is negligible. Furthermore, viewed from
low-Earth orbit, the limb is orders of magnitude brighter than the
GDE. Using the Earth's emission is, therefore, a simpler
way to derive the spectrum of CR nucleons than using the GDE.

The \gray{} emission from the Earth's limb was first observed by the SAS-2
\cite{1981JGR....86.1265T} and EGRET \cite{2005AIPC..745..709P}
instruments, but these observations were limited in statistics and
angular resolution. \fermilat{} made the first measurement of the Earth's limb
\gray{} emission above 10~GeV \cite{FermiEarth2009} and was able to resolve
the limb profile to discriminate between the thin and thick target regimes,
demonstrating its capability for indirect measurements of the CR spectrum.
In this paper we report on the analysis of 5 years of Earth's limb
observations with the \fermilat{}.

%%%%%%%%%%%%%%%%%%%%%%%%%%%%%%%%%%%%%%%%%%%%%
\emph{Data and Analysis Method.}
\fermi{} was launched in June 2008 and spent the first few weeks calibrating
the instruments during the Launch and Early Operations (L\&EO) period, during
which the LAT was tracking a few well-known bright sources, allowing the
Earth's limb to frequently enter the field of view (FoV). In September 2008,
three hours of limb-stare observations were performed. This is
the data set used in \cite{FermiEarth2009} and part of the data set in the
analysis presented here. The additional part of the data set is described as follows.

The spacecraft operates mainly in survey mode, keeping the Earth's limb far from
its boresight as it is a background for other analyses.
However, for a small fraction
of the operating time, the LAT performs pointed observations by following
a celestial target while it is not occulted by the Earth, including while
it is near the limb. We select this pointed
data set by accepting events when the magnitude of the
rocking angle\footnote{The angle between the LAT's boresight and the zenith
is called the rocking angle.} is $>52^{\circ}$, $2^{\circ}$
greater than that for the normal survey mode, up to August 8, 2013.
This rocking angle selection rejects the survey mode data, for which
the Earth's limb photons have large ($>62^{\circ}$) incidence angle.

\begin{table*}[htbp]
\begin{tabular}{cccccc}
Observation type    & Start date   & End date     & Livetime (days) & $\left<\Theta_{\rm LAT}\right>$ & N$_{>15\rm ~GeV}$ \\
\hline \hline
L\&EO               & Jul 15, 2008 & Jul 30, 2008 & 9               & $44^{\circ}$                    & 967               \\
Limb-stare          & Sep 29, 2008 & Sep 29, 2008 & 0.125           & $31^{\circ}$                    & 18                \\
Pointed (multiple)  & Aug 21, 2008 & Aug 8, 2013  & 90              & $45^{\circ}$                    & 6762              \\
\hline
\end{tabular}
\caption{Observation types and durations for this analysis.}
\label{table:datalist}
\end{table*}

We avoid the geomagnetic and solar modulation of local CRs near the
Earth by considering only $\gamma$ rays above 15~GeV because
they must be produced by CR protons with energies of at least
(but mostly much greater than) 15~GeV.
The resulting number of Earth's limb photons above 15~GeV
(N$_{>15\rm ~GeV}$) and the average incidence angles measured from the LAT's
boresight ($\left<\Theta_{\rm LAT}\right>$) for these datasets are in
Table~\ref{table:datalist}.

The data are analyzed here in the local nadir coordinates,
in which $\Theta_{\rm Nadir}$ is the angle
measured from the nadir direction at the location of the LAT.
At the LAT's altitude of $\sim 565$~km, the physical
limb of the Earth is at $\Theta_{\rm Nadir}\approx 66.7^{\circ}$.
However, the peak of the \gray{} emission above 15~GeV is at
$\Theta_{\rm Nadir}\approx 68.1^{\circ}$, due to the height of the
atmosphere and the effects of \gray{} absorption
as discussed in detail in \cite{FermiEarth2009}. At $\Theta_{\rm Nadir}
=68.4^{\circ}$, the integrated column density for grazing-incidence particles
is $\sim 3$~g~cm$^{-2}$ (see Figure~5 in \cite{FermiEarth2009}). From this
angle outwards, the atmosphere is in the thin-target regime
with photons produced from a single interaction, the absorption
effects are negligible, and the resulting \gray{} spectrum is determined by
the local spectrum of CRs. Thus,
$\Theta_{\rm Nadir}=68.4^{\circ}$ is the inner edge for the
studies presented here. The outer edge of the Earth's limb is chosen as
$\Theta_{\rm Nadir}=70.0^{\circ}$ because the emission from celestial
sources starts to dominate for larger $\Theta_{\rm Nadir}$ angles.

We use the {\texttt{P7REP}} reprocessed data (see \cite{P7REP2013} and FSSC
page\footnote{http://fermi.gsfc.nasa.gov/ssc/data/analysis/documentation/\linebreak
Pass7REP\_usage.html} for details)
with the {\texttt{P7REP\char`_SOURCE}} event selection and the associated
{\texttt{P7REP\char`_SOURCE\char`_V15}}
instrument response functions. We apply two additional cuts: 
$\Theta_{\rm LAT}<70^{\circ}$ (reduced incidence angle)
to avoid the edge of the FoV, which is prone to
systematic uncertainties, and
$68.4^{\circ}<\Theta_{\rm Nadir}<70^{\circ}$
(thin-target regime) to select photons from the Earth's limb.

\begin{figure}[]
\includegraphics[width=\linewidth]{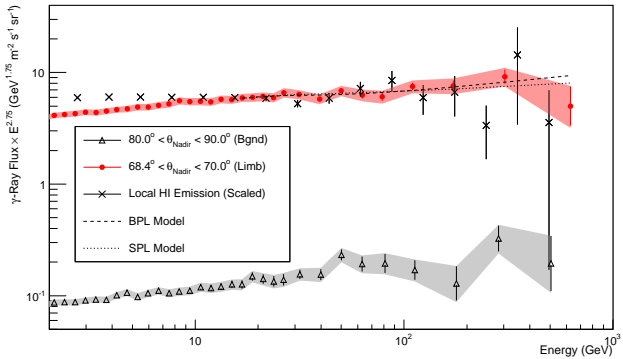}
\caption{The \gray{} energy spectra multiplied by $E^{2.75}$.
  The background-sky spectrum (triangles) has been subtracted
  from the limb spectrum (circles). The scaled local H\,{\sc i}
  emission\textsuperscript{\ref{Hi}}
  (crosses) is shown for comparison. Best-fit \gray{} results from
  two CR models based on the \emph{K\&O} model \cite{KO2012}
  are also plotted as dotted and dashed lines.
  Statistical and total (quadrature sum of statistical and
  systematic) errors are shown as bars and bands, respectively.}
\label{fig:Spectrum}
\end{figure}

The background is estimated from a ring surrounding the Earth's limb
($80^{\circ}<\Theta_{\rm Nadir}<90^{\circ}$). The ring
immediately surrounding the limb was not used in order to avoid
spill-over photons from the limb due to the LAT's point-spread function
(PSF). The background level shown in Figure~\ref{fig:Spectrum} is small,
ranging from $\sim 3$\% at 15~GeV to $\sim 5$\% at 500~GeV of the
bright limb emission.

For the {\texttt{P7REP}} data used here, dedicated
simulations and flight data comparisons have been performed to validate
the LAT responses up to 1~TeV by the LAT Collaboration. Based on these studies,
we adopt an effective area ($A_{\rm eff}$) uncertainty of 5\% at 10~GeV,
increasing linearly with the logarithm of energy to 15\% at 1~TeV\footnote{
http://fermi.gsfc.nasa.gov/ssc/data/analysis/LAT\_caveats.html}.

We simulate many realizations of the energy dependence of the
$A_{\rm eff}$ which obey the above estimated uncertainty
to observe the effect of instrumental systematic error.
Specifically, to obtain one realization, we generate 3 random numbers
at 10, 100, and 1000~GeV from Gaussian distributions for which the mean is 0 and $\sigma$
is the value of the uncertainty estimation at the three energy points. Cubic spline
interpolation between these points describes the
deviation of $A_{\rm eff}$ from the central value, which would then distort
the measured Earth's limb spectrum in a way consistent with the
systematic uncertainties, allowing us to evaluate the propagated
uncertainties of the final results. This algorithm to simulate $A_{\rm eff}$
uncertainties assumes uncorrelated errors for two energy bins that are sufficiently far apart 
(larger than half a decade in energy), but the interpolation results in highly
correlated errors between nearby energy bins (see Section 5.6.2 in \cite{P7Paper2012}).

We also correct for the angular resolution effects on the
limb spectrum itself. The dominant effect is contamination
by limb photons from $\Theta_{\rm Nadir}<68.4^{\circ}$,
where the emission is brighter. The other is the leakage 
of photons from the limb to each side of the defined boundary. Above
15~GeV, where the LAT's PSF is narrow (68\% containment at
$\sim 0.2^{\circ}$) and not strongly energy dependent
(see Figure~57 in \cite{P7Paper2012}), these corrections combined
decrease the measured intensities by $\sim 35$\%, depending on energy.
The effect on the spectral index is relatively small compared
to that from $A_{\rm eff}$.

We determined that +2\%/-5\% uncertainty
of the absolute energy scale (described in Section 7.3.4 in \cite{P7Paper2012})
translates into $<10$\% effects on the absolute normalization
of the spectrum, which does not alter the results presented here.

To infer the CR proton spectrum from the \gray{} measurement, we use two $pp\to\gamma$
interaction models, one by \citet{Kamae2006} (\emph{Kamae} model) and the other
by \citet{KO2012} (\emph{K\&O} model). For each model, we calculate
the \gray{} spectrum by integrating a model of the proton spectrum from $\sim 0.5$~GeV
to $\sim 500$~TeV in kinetic energy.

In our study, we assume that the atmosphere consists of 100\% Nitrogen.
This does not affect our results because studies of proton-nucleus interactions
at high energies (e.g., \cite{Orth1976,Atwater1986})
show that the $pA$ cross section can be scaled from the $pp$ cross section
by applying an energy-independent scaling factor $\propto$$A^{0.7}$,
where $A$ is the atomic number of the nucleus. The precise scaling factor
for the atmospheric composition is not important for this analysis because it
changes only the normalization of the fitted proton spectrum.

Observations of the \gray\ emission from the Earth's limb cannot discriminate
between contributions of CR protons and heavier nuclei. Thus, we must rely on
the direct measurements of the CR composition.
The He fraction in CRs is about 6--10\% by number, depending on energy,
in our energy range of interest, so its contribution has to be
taken into account, while the contribution of the heavier nuclei can be safely neglected.
There are a number of empirical parameterizations
for nucleus-nucleus meson multiplicity (e.g., Appendix A in \cite{Orth1976} and
Eq.~(3b) in \cite{Atwater1986}). These formulas give similar values for
the ratio of $\alpha N$ to $p N$ cross sections, $\sigma_{\alpha N}$/$\sigma_{p N} \sim 1.6$.
We use this number to scale the $pp$-interaction models and to calculate the relative
contribution of He nuclei to the limb \gray{} emission. Since the contribution of
$\gamma$ rays produced by He is $\sim 10$--20\% depending on energy, and most of the emission
is produced by protons, the $\alpha N$ scaling uncertainty has little influence on the final fit
results.

To determine the He spectrum, we fit the combined PAMELA
(\cite{PAMELA2011}), CREAM (\cite{CREAM2011pHe}), and ATIC-2 (\cite{Panov2009})
He data above 50~GeV/n (102~GV) with spectral forms described
below Equation~\ref{eq:LH}. 
According to PAMELA measurements \cite{PAMELA2011}, the He/$p$ ratio at $\sim 90$~GeV
is 6.2\%. We use this value together with the cross section scaling to fix the
contribution of He to photon production at 15~GeV to 9.5\%.
We then forward-fold by varying the parameters
of the input proton spectrum so that the resulting \gray{} spectrum calculated from
the $pp$-interaction models provides the best fit to the Earth's limb measurement.
In the fitting procedure, the normal Poisson likelihood function is maximized:

\begin{equation}
L=\prod_{i=1}^{N}P_{\rm Poisson}(n_i^{\rm obs},n_i^{\rm mod}),
\label{eq:LH}
\end{equation}
where $N$ is the number of energy bins. $P_{\rm Poisson}$ is the Poisson probability of
observing $n_i^{\rm obs}$ counts given that
the model predicts $n_i^{\rm mod}$ counts (\gray{} model flux $\times$ exposure)
for the $i^{\rm th}$ energy bin. We use two models for the local CR proton and He
spectra in the fitting procedure:

\begin{itemize}
\item SPL: Single power law in rigidity. This model assumes a single power
 law for CR protons and He. For He, the index $2.73 \pm 0.01$
 is our best-fit value of the combined PAMELA, CREAM,
 and ATIC-2 data. For protons, we fit both the normalization and index 
 to our measurement of the \gray{} spectrum from the Earth's limb.

\item BPL: Broken power law in rigidity. This model assumes a broken power
 law for both proton and He spectra. As before, the He
 spectrum is fixed to the best fit of the combined direct measurements,
 for which the spectral index changes from $2.82 \pm 0.07$ to
 $2.55 \pm 0.02$ at $247 \pm 44$~GV.
 We then fit the indices, break energy ($E_{\rm break}$), and normalization
 for the proton spectrum.
\end{itemize}

We evaluate the statistical uncertainties of the fit results by
fitting a large number of simulated realizations of photon counts generated
with a Poisson distribution for which the expected value is the measured count
in each energy bin.
Likewise, simultaneous simulations of photon counts and ranges of $A_{\rm eff}$,
as previously described, give the total (combined systematic and statistical)
errors. We also add in quadrature the 5\% absolute energy scale uncertainty
to the errors of the fitted energy parameter.

%%%%%%%%%%%%%%%%%%%%%%%%%%%%%%%%%%%%%%%%%%%%%
\emph{Results.}
The measured \gray{} thin-target limb spectrum, the background-sky
flux (which has already been subtracted from the limb spectrum),
and the best-fit \gray{} models are shown in Figure~\ref{fig:Spectrum}.
The \gray{} emission from the local H\,{\sc i}\footnote{Work in
preparation by the LAT Collaboration \label{Hi}} is scaled to approximately
match that from the Earth's limb and shown for comparison. As
expected, the two agree well above $\sim 10$~GeV, since they are
produced from hadronic interactions by the same
local population of CRs in the thin-target regime. Below $\sim 10$~GeV,
the spectra differ due to the geomagnetic and solar modulations of CRs
in the vicinity of the Earth, reducing the number of CRs interacting
with the Earth's atmosphere. For this reason we limited our
study to limb $\gamma$ rays above 15~GeV. The
approximate proton-to-\gray{} energy conversion factor for the power-law
spectrum of protons is 0.17 \cite{Kelner2006}. The energy range
of the inferred proton spectrum is thus $\sim 90$~GeV--6~TeV.

\begin{figure*}[]
\includegraphics[width=0.9\linewidth]{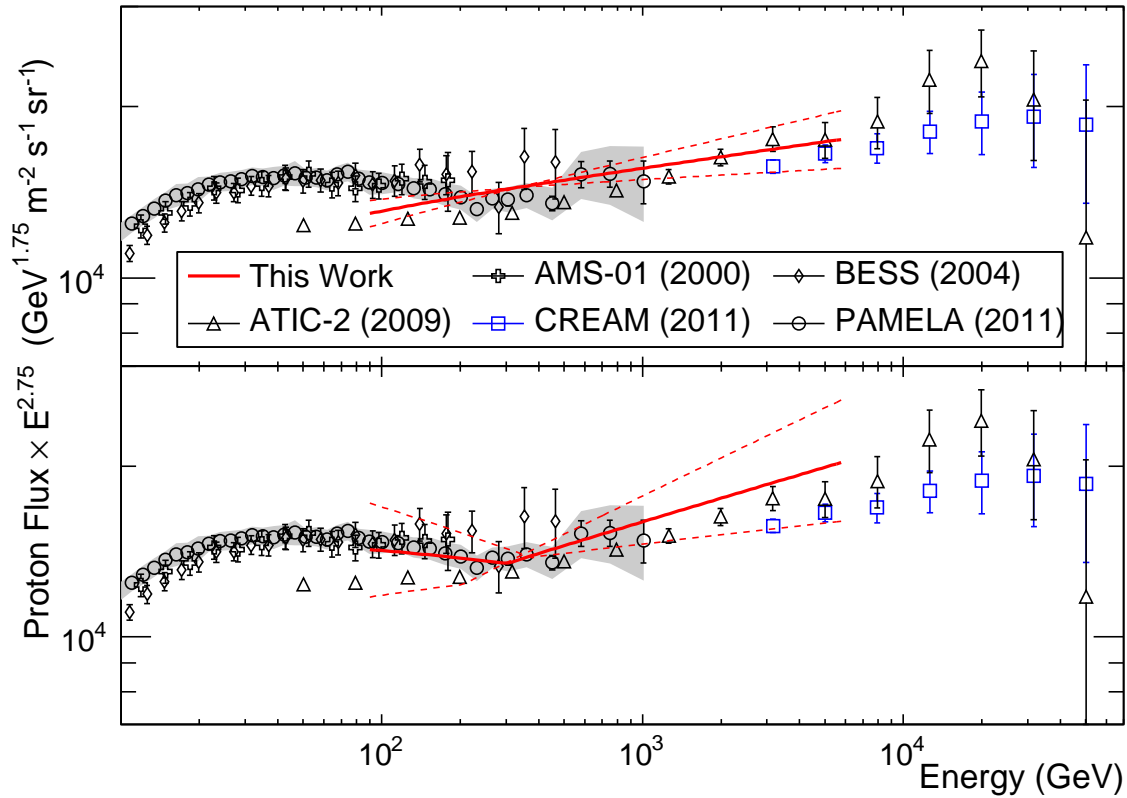}
\caption{Best-fit single power law
  or SPL (top), and best-fit broken power law or BPL (bottom)
  spectrum for the local CR proton spectrum (solid red lines) as derived from
  the Earth's limb \gray{} data using the \emph{K\&O} model \cite{KO2012} for $pp$-interactions.
  The total (combined statistical and systematic, neglecting errors in absolute normalization)
  uncertainties are the dashed red lines.
  Other direct measurements (\cite{Panov2009},\cite{CREAM2011pHe},\cite{PAMELA2011},\cite{AMS2000},\cite{BESS2004})
  are shown for comparison. The gray band is PAMELA's total uncertainty.}
\label{fig:Proton}
\end{figure*}

\begin{table}[h!]
\begin{tabular}{ccc}
                    & \emph{Kamae}           & \emph{K\&O}            \\
\hline \hline
SPL index           & $2.67 \pm 0.05~(0.03)$ & $2.68 \pm 0.04~(0.03)$ \\
BPL index 1         & $2.81 \pm 0.10~(0.03)$ & $2.81 \pm 0.11~(0.04)$ \\
BPL index 2         & $2.60 \pm 0.08~(0.05)$ & $2.61 \pm 0.08~(0.06)$ \\
BPL $E_{\rm break}$ & $276 \pm 64~(55)$~GeV  & $302 \pm 96~(62)$~GeV  \\
BPL vs SPL          & $1.0~\sigma$           & $1.0~\sigma$           \\
\hline
\end{tabular}
\caption{Fit results from \emph{Kamae} \cite{Kamae2006} and \emph{K\&O} \cite{KO2012}
  models, shown as $value~\pm~total~error~(statistical~error)$.}
\label{table:fitresult}
\end{table}

Using the \emph{K\&O} and \emph{Kamae} models, we obtain the results
shown in Table~\ref{table:fitresult}. The log likelihood for the best-fit BPL
is $\sim 0.9$ better than that for the best-fit SPL. To account for
$A_{\rm eff}$ systematic uncertainties,
we apply Monte Carlo simulations to translate this likelihood ratio
into a significance. By assuming that the best-fit SPL is the true underlying
flux model, we produce $\sim 2000$ simulations of $A_{\rm eff}$ from the
estimated errors as previously discussed, for each of which we generate
$\sim 1000$ simulations of SPL realizations. We then fit the distribution of the
log likelihood differences between SPL and BPL for these $\sim 2$M
total simulations with a Gaussian function and evaluate how likely it is that
the best-fit BPL we obtain from the actual measurement
would give a log likelihood difference of 0.9 or above as compared to the SPL.
We find that it corresponds to a significance of $1.0~\sigma$.

We performed several cross checks to test the stability and consistency
of the results. We studied the effects of using
the event selection with more stringent rejection of residual CRs
({\texttt{P7REP\char`_CLEAN}}), tighter incidence angle ($\Theta_{\rm LAT}$) cuts, and
reasonable variations of the fitted energy ranges (up to 20~GeV lower bound and
down to 120~GeV upper bound in \gray{} energy). All of these cases yield
consistent results.

Figure~\ref{fig:Proton} shows the resulting best-fit SPL and BPL derived from the \emph{K\&O}
model in comparison with direct measurements, assuming an effective
atmospheric column density of $\sim 1.0$~g~cm$^{-2}$, as described below.

In order to determine the absolute normalization of the inferred proton spectrum,
we use the NRLMSISE-00 atmospheric model \cite{Picone2002} to calculate the
average line-of-sight column density, weighted by \gray{} intensity,
in the range studied here ($\Theta_{\rm Nadir}=68.40^{\circ}$--$70.00^{\circ}$) to be
1.2~g~cm$^{-2}$. Due to the exponential change of the atmospheric density
with $\Theta_{\rm Nadir}$, the evaluated density is extremely sensitive
to the lower bound of the $\Theta_{\rm Nadir}$ range.
We thus empirically adjust the absolute normalization of our inferred proton
spectrum to approximately match that of direct measurements as shown in
Figure~\ref{fig:Proton} by changing the atmospheric column density from
1.2~g~cm$^{-2}$ to 1.0~g~cm$^{-2}$. This is equivalent to increasing the
lower bound of $\Theta_{\rm Nadir}$ from 68.40$^{\circ}$ to 68.42$^{\circ}$
when we calculate the atmospheric column density. The small change in the
effective lower bound of $\Theta_{\rm Nadir}$ by $\sim 0.02^{\circ}$ has many 
potential justifications, such as the LAT altitude variations which smear the
precise calculation of the target density, the atmospheric model uncertainties,
and other absolute normalization uncertainties as previously discussed.
Since our primary interest is in the spectral indices, the difference
in normalization is of no importance.

%%%%%%%%%%%%%%%%%%%%%%%%%%%%%%%%%%%%%%%%%%%%%
\emph{Discussion and Conclusion.}
Our LAT analysis, which employs a different technique from
direct measurements, shows that the CR proton spectrum between
$\sim 90$~GeV--6~TeV
can be described equally well ($\sim 1~\sigma$) with the SPL and BPL models.
The best-fit spectral indices ($2.68 \pm 0.04$ for SPL and $2.61 \pm 0.08$
above $\sim 200$~GeV for BPL) are consistent with each other.

We note that our best-fit SPL index is $\sim 3~\sigma$ from the
value ($2.801 \pm 0.007$) reported by PAMELA for a lower energy range
(29--79~GeV). However, our best-fit SPL index for $\sim 90$~GeV--6~TeV is in
good agreement with the fitted index for $\sim 230$~GeV--1~TeV reported by
PAMELA \cite{PAMELA2011} and with the measurements at higher
energies by ATIC-2 \cite{Panov2009} and CREAM \cite{CREAM2011pHe}.
While \fermilat{} results cannot confirm or disprove the existence of the
spectral break itself yet, they do indicate a flatter proton spectrum
at high energies, consistent with direct measurements by ATIC-2 and CREAM.

This result is the first indirect measurement of the proton spectrum
in the energy range $\sim 90$~GeV--6~TeV
using observations of the \gray{} emission of the Earth's limb. Continuing
observations with \fermilat{} will allow us to improve the precision of the
measurement of the CR spectrum and extend the energy range.

The \fermilat{} Collaboration acknowledges support from a number of
agencies and institutes for both development and the operation of the
LAT as well as scientific data analysis. These include NASA and DOE in
the United States, CEA/Irfu and IN2P3/CNRS in France, ASI and INFN in
Italy, MEXT, KEK, and JAXA in Japan, and the K.~A.~Wallenberg
Foundation, the Swedish Research Council and the National Space Board
in Sweden. Additional support from INAF in Italy and CNES in France
for science analysis during the operations phase is gratefully
acknowledged. I.V.M.\ acknowledges support from NASA grants
NNX11AQ06G and NNX13AC47G.

%%%%%%%%%%%%%%%%%%%%%%%%%%%%%%%%%%%%%%%%%%%%%

\bibliography{Biblio}

\end{document}